# The Digital Agricultural Revolution: A Bibliometric Analysis Literature Review

RICCARDO BERTOGLIO[1], CHIARA CORBO[2], FILIPPO M. RENGA[2],
AND MATTEO MATTEUCCI[1], (Member, IEEE)
[1]Department of Electronics, Information and Bioengineering, Politecnico di Milano, 20133 Milan, Italy
[2]Department of Management, Economics and Industrial Engineering, Politecnico di Milano, 20156 Milan, Italy

Corresponding author: Riccardo Bertoglio (riccardo.bertoglio@polimi.it)

**ABSTRACT** The application of digital technologies in agriculture can improve traditional practices to adapt to climate change, reduce Greenhouse Gases (GHG) emissions, and promote a sustainable intensification for food security. Some authors argued that we are experiencing a Digital Agricultural Revolution (DAR) that will boost sustainable farming. This study aims to find evidence of the ongoing DAR process and clarify its roots, what it means, and where it is heading. We investigated the scientific literature with bibliometric analysis tools to produce an objective and reproducible literature review. We retrieved 4995 articles by querying the Web of Science database in the timespan 2012-2019, and we analyzed the obtained dataset to answer three specific research questions: i) what is the spectrum of the DAR-related terminology?; ii) what are the key articles and the most influential journals, institutions, and countries?; iii) what are the main research streams and the emerging topics? By grouping the authors' keywords reported on publications, we identified five main research streams: Climate-Smart Agriculture (CSA), Site-Specific Management (SSM), Remote Sensing (RS), Internet of Things (IoT), and Artificial Intelligence (AI). To provide a broad overview of each of these topics, we analyzed relevant review articles, and we present here the main achievements and the ongoing challenges. Finally, we showed the trending topics of the last three years (2017, 2018, 2019).

## NOMENCLATURE

| | |
|---|---|
| AI | Artificial Intelligence |
| CSA | Climate-Smart Agriculture |
| DAR | Digital Agricultural Revolution |
| FAO | Food and Agriculture Organization |
| FMIS | Farm Management Information System |
| GC | Global Citations |
| GHG | Greenhouse Gases |
| GIS | Geographic Information System |
| GNSS | Global Navigation Satellite System |
| IF | Impact Factor |
| IoT | Internet of Things |
| JCR | Journal Citation Reports |
| LC | Local Citations |
| ML | Machine Learning |
| NDVI | Normalized Difference Vegetation Index |
| PA | Precision Agriculture |
| PLF | Precision Livestock Farming |
| RS | Remote Sensing |
| SCR | Standard Competition Ranking |
| SSM | Site-Specific Management |
| UAS | Unmanned Aerial System |
| UAV | Unmanned Aerial Vehicle |
| UGV | Unmanned Ground Vehicle |
| VRT | Variable Rate Technology |
| WoS | Web of Science |
| WSN | Wireless Sensor Network |

## I. INTRODUCTION

A worldwide ever-growing population has to cope with the limited resources of our planet. The Food and Agriculture

  

Organization (FAO) stated that in 2050 we would be 9 billion people, and the food demand will grow by 70% [1]. However, arable land is limited, and climate change endangers crop yields. Reacting to these threats is of paramount importance. Indeed, the UN 2030 agenda within its 17 Sustainable Development Goals (SDGs) has planned, among other objectives, to reach sustainable food production systems via agricultural practices that increase productivity and that adapt to climate change [2].

Reaching more productive and sustainable agriculture within a changing climate is a tough challenge; agricultural practices need to be revolutionized to meet the SDGs by 2030. The ongoing process of agricultural digitalization seems promising to face the upcoming challenges. This revolutionary process is called DAR and is bringing innovations to support the farmers by increasing crop yields while reducing the environmental impact [3]. In 2019, the FAO stated that "market forecasts for the next decade suggest a "digital agricultural revolution" will be the newest shift which could help ensure agriculture meets the needs of the global population into the future" [4].

To respond to the challenges mentioned above, international institutions like the World Bank and the FAO have recommended a global transition to the CSA framework. The CSA framework can be defined as "a strategy to address the challenges of climate change and food security by sustainably increasing productivity, bolstering resilience, reducing GHG emissions, and enhancing achievement of national security and development goals" [5]. Since its origin in 2007, the CSA concept has appeared in many forms. The advent of digital technologies in agriculture has shaped CSA giving birth to new terms like Smart Agriculture, Digital Agriculture, and Agriculture 4.0.

Digital technologies like AI, Robotics, and the IoT are expected to be game-changers to achieve the CSA objectives. Digital technologies allow for detailed real-time data analysis. Data are collected by smart sensors, ground robots (Unmanned Ground Vehicle (UGV)), aerial drones (Unmanned Aerial System (UAS)), or satellites [6]. This big amount of data is analyzed by AI algorithms producing information. Thus, farmers can make decisions both by exploiting their in-field experience and data analysis. Digital technologies have the potential of increasing productivity while decreasing costs and being more environmentally friendly.

The DAR is a growing trend in the research community and the private sector. However, there is no general agreement about when the DAR started or about its enabling technologies. In the scientific literature, authors refer to the DAR in different ways. Many authors use the term Precision Agriculture (PA), introduced in the '90s [7], [8]. Others prefer the more modern terms "Smart Agriculture" or "Digital Agriculture" [9], [10]. Most recently, the use of Agriculture 4.0 is becoming commonplace [11], [12].

In this work, we relied on bibliometrics tools to give a complete description of the DAR research field's boundaries and dynamics. Bibliometrics is applying quantitative analysis and statistics to published articles to measure their impact [13]. Traditionally, the two main methods for synthesizing past research findings have been the qualitative approach of a structured literature review and the quantitative approach of meta-analysis. While the qualitative approach suffers from the subjective biases of the researchers involved in the process [14], bibliometrics is a method to perform a systematic, objective and reproducible analysis of the scientific activity. This is particularly useful in a fragmented domain such as that of DAR [15].

Literature reviews are increasingly becoming important due to the fast-growing pace of scientific production. Synthesizing the existing knowledge base is crucial for advancing the line of research. Bibliometric methods become particularly useful to provide an objective analysis of an unstructured and large body of information. They can be used to detect the emerging and trending topics, the most influential journals/institutions/countries, the principal research streams, and to show the "big picture" of a research field [15].

According to our best knowledge, there are still no literature reviews that exploited bibliometric analysis for summarizing the entire DAR research field. Existing bibliometric studies are based on incomplete queries that do not account for word combinations or lack relevant terms. For instance, the authors of [16] made a bibliometric analysis review of information and communication technologies (ICT) in agriculture. However, they limited their research to a subset of the literature. Also, they did not perform a data preparation and cleaning phase, which is crucial for identifying relevant groups of words (topics). Similarly, the authors of [7], [8], focused their query only on the PA field. Other bibliometric analysis works are focused on more specific themes: in [17], the authors did a review about Big Data in agriculture; in [18] the authors made a bibliometric analysis regarding the digital technologies for plant phenotyping; in [19] they did a bibliometric literature review about the digital agricultural techniques in the coffee sector; in [20] the authors focused on the use of UAS in agriculture.

This paper aims to a comprehensive analysis of the DAR research field with the tools of bibliometrics to answer three specific research questions:

1) *What is the spectrum of the DAR-related terminology?*
2) *What are the key articles and the most influential journals, institutions, and countries?*
3) *What are the main research streams and the emerging topics?*

By answering the above research question, this paper aims to provide researchers a comprehensive and objective view of the DAR research field. It also supports the research process by identifying potential collaborations, the key journals to publish on, and giving research directions by identifying the emerging topics. This paper is also helpful for policy-makers to have a complete view of the possibilities, advantages, and challenges associated with the DAR.

The outline of the paper is as follows. In Section II, we present an overview of the DAR related terminology.

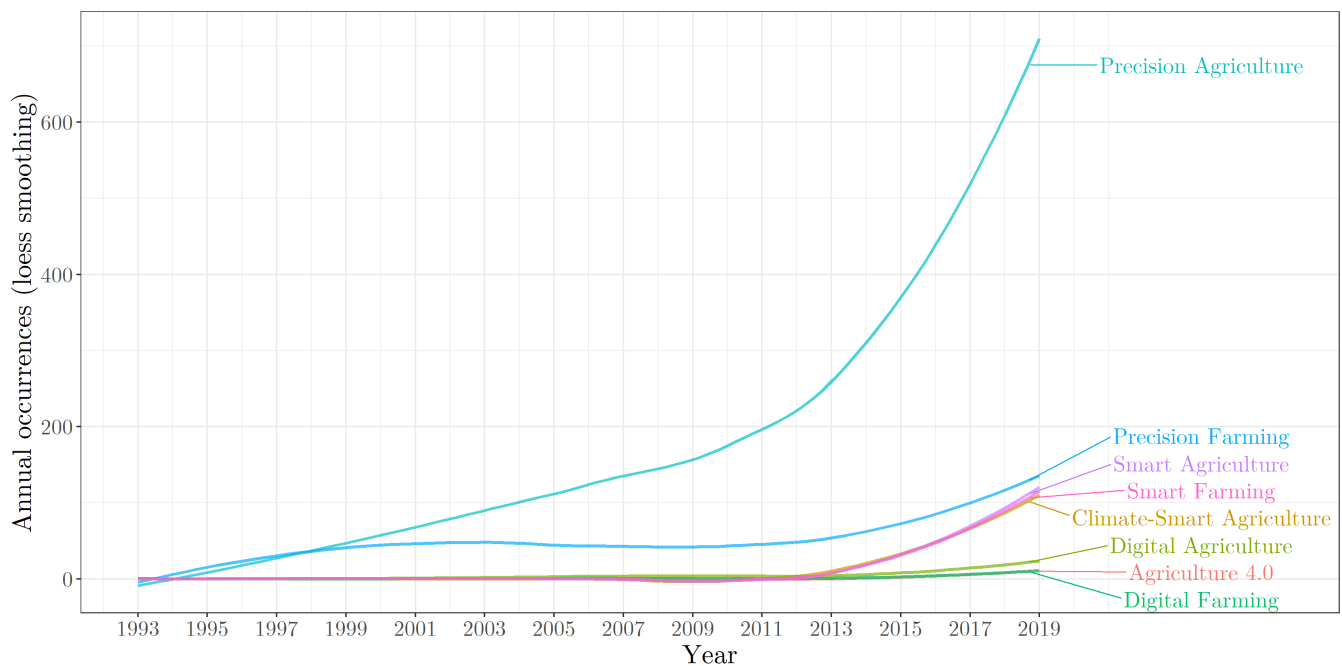

**FIGURE 1.** Yearly smoothed number of studies published in the WoS database from 1993 to 2019 containing the corresponding term.

In particular, we focused our analysis on the recently emerging "Agriculture 4.0" term. We summarized the different authors' opinions about its definition and origin. In Section III, we present the methodology and the tools used for the bibliometric analysis. We motivated the choice of the publication database and clarified the data preparation process. In Section IV, we present the bibliometric analysis results along with our interpretation. We grouped the author's keywords to obtain the main research streams, and for each one, we selected and presented some relevant review articles. Finally, in Section V, we provide research directions and draw some conclusions.

## II. DAR TERMINOLOGY AND BOUNDARIES

PA, CSA, Smart Farming, Digital Agriculture, and Agriculture 4.0 are only part of the names used to refer to the DAR. To make some order, we analyzed the literature employing the Web of Science (WoS) database. We aimed at figuring out how these terms' popularity evolved over the years.

We retrieved from the WoS database the yearly number of studies in the time-span 1993-2019 (being 1993 the year of the oldest publication in the database) that contained one or more of the following terms: "precision agriculture", "precision farming", "climate-smart agriculture", "smart agriculture", "smart farming", "intelligent agriculture", "intelligent farming", "digital agriculture", "digital farming", "agriculture 4.0".

Figure 1 shows the year-wise smoothed distributions of the number of studies published between 1993-2019 that contain each of the terms above. We smoothed the year-wise distributions by using the R *loess()* function that is based on Local regression, a kind of non-parametric regression model [21]. In the figure, note that the distribution of the "Digital Farming" term overlaps that of "Agriculture 4.0". Also, the terms "Intelligent Agriculture" and "Intelligent Farming" have been excluded from the following analysis since they showed similar year-wise distributions (even if of a smaller entity) to those of the terms "Smart Agriculture" and "Smart Farming", respectively. Indeed, "intelligent" and "smart" are synonyms, being the latter more used in the DAR research field.

From Figure 1 it is possible to see that the term PA is the most used in the literature. PA, "or information-based management of agricultural production systems, emerged in the mid-1980s as a way to apply the right treatment in the right place at the right time" [22]. PA has been enabled by technologies such as Global Navigation Satellite System (GNSS), Geographic Information System (GIS), and microcomputers. These technologies led to innovations like Variable Rate Technology (VRT) or the auto-guidance systems.

VRT consists in spreading a variable quantity of resources (water, fertilizers, and pesticides) by considering the soil and crop variability across the field. The auto-guidance systems exploit GNSS signal and fields' maps to compute smart routes for agricultural machines. Routes are computed to minimize multiple coverages of the same field area, a well-known agricultural problem that causes waste of resources.

The first appearance of the term PA in the WoS database dates back to 1994. However, technology is now only advanced enough to realize the PA concepts in practice. Thus, digital technologies are supporting the application of PA practices. In the scientific literature, PA can also be found with its synonym Precision Farming.

In Figure 1 it is also possible to recognize a set of trending terms: "Climate-Smart Agriculture", "Smart Agriculture", and "Smart Farming". They all saw a rapid increase since 2012, two years after the formalization of the concept of CSA from the FAO in 2010 [23]. CSA is one of the five main research streams that we identified in this study, and it will be treated in depth in Section IV-D1.

The temporal evolution of the terms "Climate-Smart Agriculture", "Smart Agriculture", and "Smart Farming" suggests a connection between them since they show a similar year-wise distribution. We could not find review articles with a broad perspective on the Smart Agriculture term in the scientific literature. Thus, we relied on articles treating specific applications to extract the authors' definitions of the Smart Agriculture term. From these definitions, it can be stated that Smart Agriculture shares the same objectives of CSA while having a focus on digital technologies.

CSA it is a general framework for promoting sustainable agricultural practices that adapt to climate change. However, CSA does not specify which are the means for reaching its objectives. Conversely, the entirety of analyzed studies regarding "Smart Agriculture" are based on digital technologies [9]–[12], [24]–[27]. "Smart Agriculture" (or "Smart Farming") can be defined as the application of digital technologies for reaching the CSA objectives.

The term "Digital Agriculture" appeared for the first time in the WoS database in 2002. Since then, it has only gained relatively little attention in recent years. From an inspection of the literature, it can be stated that the terms "Digital Agriculture" and "Digital Farming" are neither more than synonyms of "Smart Agriculture". Again, we could not find review articles with a broad perspective on the Digital Agriculture term, and thus we relied on the definitions found in specific studies. From an inspection of the literature, it can be stated that Digital Agriculture, similarly to Smart Agriculture, shares the same objectives of CSA while showing a focus on digital technologies [12], [28]–[30].

Finally, the emerging term "Agriculture 4.0" has appeared for the first time in the WoS database in 2016. Even if this term's popularity is still in its infancy, the FAO formally adopted it for the first time in 2020 [31]. Thus, it is foreseeable that "Agriculture 4.0" will gain more attention in future years, with a diffusion similar to that of CSA after its formalization in 2010. Following this reasoning, we investigated the scientific literature to clarify the brand-new "Agriculture 4.0" concept.

### A. WHAT IS AGRICULTURE 4.0?

In November 2020, the FAO published an article about Agriculture 4.0 [31] describing it as "agriculture that integrates a series of innovations in order to produce agricultural products. These innovations englobe precision farming, IoT and big data in order to achieve greater production efficiency".

However, a commonly accepted definition of the term "Agriculture 4.0" still does not exist. To give the reader a context and the boundaries of the emerging Agriculture 4.0 concept, we analyzed studies that used this term. We found 45 studies by querying all the WoS database's indices on the 26th of November 2020. We used "agriculture 4.0" as the query "Topic" without any filter. Of these 45 studies, 13 did not explain the term "Agriculture 4.0", 7 were not accessible by us or duplicated. Thus, we analyzed the remaining 25 studies and collected the explanations the authors gave of Agriculture 4.0. By summarizing all these explanations, in the following, we report a description of Agriculture 4.0.

Agriculture 4.0 is a neologism coming from the concept of Industry 4.0. It is associated with the Digital Agricultural Revolution since its pillars are digital technologies such as AI, Big Data, Cloud Computing, Robotics, IoT, RS, and the Blockchain [26], [32]. The aim of Agriculture 4.0 is of optimizing agricultural tasks by reducing inputs (water, fertilizers, pesticides) [33] and increasing farms productivity coping with climate change [26], [34]. Agriculture 4.0 provides tools to increase the profit margin for farmers and reduce the risk of environmental contamination.

According to Agriculture 4.0, farmers are assisted by a Decision Support System that guides them in programming the treatments [26], [35]–[38]. They will no longer have to apply water, fertilizers, and pesticides uniformly across entire fields. On the contrary, farmers will use the minimum quantities of resources that the plants require. VRT [39] enables for precise and targeted spraying of substances with a centimeter accuracy [40], [41].

Data are collected in a multi-sensorial way by operating at different scales, spatial (where) and temporal (when). Regarding imaging sensors, there is another scale to consider, that is, spectral (what). Indeed, imaging data can be collected at different light wavelengths [39]. Sensors are installed in the fields and on robotics platforms like UGV or Unmanned Aerial Vehicle (UAV). Satellites data are also employed, and multiple-source data integration is performed [35].

Agriculture 4.0 stands for the combined internal and external integration of farming operations [35], [42]. Prediction models help to handle better external factors such as weather conditions, market behaviors, and current trends in needs [43]. Moreover, information is collected along the entire supply chain for integration and traceability purposes [34], [44].

Agriculture 4.0 envisages the entire value chain, from the farmer to the distribution, connected via the internet to coordinate and share information. The value chain's actors are connected with a continuous and data-rich communication thanks to the virtualization of the processes [42]. Physical and virtual objects can interconnect and interact autonomously [37].

Agriculture 4.0, thanks to wideband connection technology and cloud storage, allows a real-time view on the farm and creates agro-ecosystems of connected farms and machinery [45], [46]. Agriculture 4.0 brings a new concept of whole-farm management based on the cross-industry cooperation of stakeholders, infrastructures, and technologies [37]. Thus, all the involved actors can decide even on issues outside their expertise area [47].

Finally, some authors include in Agriculture 4.0 advances from other disciplines beyond those related to digital technologies. These advances are genetic engineering, 3D-printing-based food supply, meat culturing, vertical farming, aquaponics, and circular agriculture [36], [47], [48].

### B. ORIGINS OF AGRICULTURE 4.0

It is still not completely clear where the term Agriculture 4.0 comes from. It is explained by many authors as an analogy to the Industry 4.0 revolution [19], [33], [35], [37], [42], [43], [45], [46], [48]–[50]. Indeed, the Industry 4.0 revolution has been characterized by the automation of traditional industrial practices by the integration of advanced digital technologies like the IoT and processes virtualization.

Other authors say that Agriculture 4.0 is a fourth of a sequence of agricultural revolutions and eras. However, there are multiple schools of thought (SoT) about this historical view:

- SoT1 - The Agriculture 1.0 era was labor-intensive and characterized by animal forces. Simple tools were used in agricultural activities. The Agriculture 2.0 era originated with the mechanization process brought by the combustion engine. Farmworks significantly increased in productivity and efficiency. The Agriculture 3.0 era was that of PA, starting when military GPS signals were made accessible for public use. PA helped reduce chemicals through innovations like the VRT [26], [32], [42].
- SoT2 - The Agriculture 1.0 era was characterized by motorization. The Agriculture 2.0 era originated from mechanization. The Agriculture 3.0 era was driven by humanism and electronics [50].
- SoT3 - The first agricultural revolution was characterized by mechanization (Agriculture 1.0), the second was constituted by the Green Revolution and its genetic modifications (Agriculture 2.0), and the third agricultural revolution was PA (Agriculture 3.0) [51].
- SoT4 - The first revolution is the transition from hunter-gathers to settled agriculture (Agriculture 1.0). The second originated from the innovations of the British Agricultural Revolution (Agriculture 2.0), and the third was characterized by the Green Revolution (Agriculture 3.0) [11], [48].

SoT1 seems to be the most supported by researchers based on the number of studies that reported it. However, it is difficult to determine which SoT is correct since naming an event as a revolution is subjective. Also, agriculture did not evolve in the same way in every part of the globe. Thus, every SoT can be considered correct apart from SoT4. In this case, the transition from hunter-gathers to settled agriculture cannot be considered an agricultural revolution since it is, in fact, the birth of agriculture. One of the proposed SoT can be used to explain the origin of the Agriculture 4.0 term. However, such plurality of different hypotheses suggests that the term "Agriculture 4.0" is more likely an analogy of Industry 4.0 than the fourth in a series of agricultural revolutions.

### C. ETYMOLOGY OF DAR

The term PA first appeared in the literature in the 1990s, and it is today the most used to refer to the DAR as shown by Figure 1. In the same Figure, it is also possible to note a clear rising trend of three terms: CSA, Smart Agriculture, and Smart Farming. From an inspection of the literature, it is possible to conclude that Smart Agriculture (and Smart Farming) is applying digital technologies for reaching CSA objectives. An inspection of the literature suggests that Digital Agriculture (or Digital Farming) is used interchangeably with Smart Agriculture without showing any relevant difference.

The term "Agriculture 4.0" has been described by an FAO's publication for the first time in 2020. However, more contributions are needed to define its definition and boundaries clearly. The literature analysis presented in Section II-A shows that Agriculture 4.0 is similar to the concept of Smart Agriculture while stressing the beyond-farm information usage.

From the description we gave in Section II-A, we can observe that some authors include in Agriculture 4.0 innovations from other disciplines beyond those related to digital innovations. Genetic engineering, 3D-printing-based food supply, meat culturing, vertical farming, aquaponics, and circular agriculture are all considered part of Agriculture 4.0. In this perspective, Agriculture 4.0 contains but is not limited to the DAR. Thus, Agriculture 4.0 could describe a broader revolution characterized by digital technologies and disruptive innovations in many fields.

In conclusion, the CSA framework gives a purpose to the DAR. The DAR is most commonly called PA, but terms like Smart Agriculture, Smart Farming, Digital Agriculture, and Digital Farming are becoming commonplace. Agriculture 4.0 is the most recently appeared term. More investigation is needed to understand if it affirms a new technological and paradigm shift or just a marketing buzzword reflecting a new fashion.

## III. MATERIALS AND METHODS

Despite naming conventions, it is undoubted that digital technologies are revolutionizing traditional agricultural practices in a process that we called DAR. The number of contributions is rapidly increasing over the years. In 2019, studies using the term PA reached an impressive number of 784 publications. At such a publication pace, it is not easy to keep track of the significant advances in the field. Systematic literature reviews are limited by the human capability of reading and analyzing studies.

In this framework, bibliometric analysis tools can help speed up the review process by quantitatively analyzing publications data. Also, by analyzing a comprehensive set of publications, bibliometrics helps scientists not miss research streams outside their actual knowledge. In the DAR research field, authors use many different terms to refer to the same concepts. The bibliometric analysis becomes particularly useful in such a fragmented field.

We adopted bibliometrics tools to discover the key articles, the most influential sources, institutions, and countries. We also aimed at highlighting the main research streams and presenting the main achievements and ongoing challenges. Finally, we presented the trending topics.

### A. MATERIALS

To perform the bibliometric analysis, we used the open-source *bibliometrix* R package (version 3.0.4) [15]. The *bibliometrix* package allows researchers to import a publications dataset and convert it into R format. The package contains algorithms for analyzing publications datasets with bibliometrics techniques. *bibliometrix* allows performing co-citation, coupling, collaboration, co-word, and network analysis. The unit element of the dataset is a publication. Each publication has associated several variables: authors, document title, document type, authors' keywords, source (publisher name), cited references, year, times cited, author address, and others.

### B. METHODS

A typical bibliometric analysis workflow is composed of several steps [14]:
- Study design: research questions are defined, and the proper time span is selected
- Data collection: the publications database is chosen, and a proper search query is formulated
- Data preparation: the dataset is cleaned and pre-processed for being analyzed
- Data analysis: the data selected are analyzed with bibliometric tools
- Data visualization: the proper visualization method is chosen
- Interpretation: the visualized results are interpreted

In the following, we enter in more detail about each of the steps to better understand the choices we have made in our bibliometrics analysis.

#### 1) STUDY DESIGN

We reported the research questions of this works in Section I. The time span chosen is 2012-2019. We chose the year 2012 as the beginning of the time span because we noticed a clear growth in usage of DAR related terms since 2012 (Figure 1). These terms are CSA, Smart Agriculture, and Smart Farming. In Figure 1 it is also possible to recognize a faster-rising trend of the terms PA and Precision Farming around the year 2012. We provide other reasons for the year 2012 as the beginning of the time span in Section IV. 2019 was the last year for retrieving complete bibliographic data when working on this article.

#### 2) DATA COLLECTION

There exist many sources of publication data: WoS, Scopus, Dimensions, Crossref, Microsoft Academic, and Google Scholar are the most common. Publications databases can be classified along a one-dimensional line that indicates the intensity of selection policy applied by the database's Content Selection Board. In one direction, they lie databases more focused on selectivity, while in the opposite direction, they lie databases more focused on comprehensiveness.

The authors of [52] made a large-scale comparison of the five most common publications databases, namely, Scopus, WoS, Dimensions, Crossref, and Microsoft Academic. They did not include Google Scholar for difficulties in retrieving large-scale data. Using the results reported in [52], we plotted the publications databases on a one-dimensional line in Figure 2. We took a database's total number of documents as a proxy of its selection policies' intensity. Web Of Science showed to be the most selective while Microsoft Academic being the more comprehensive. Scopus showed to be still focused on selectivity but with broader coverage than Web of Science.

WoS and Scopus are the two most used publication databases. Indeed, in the past, there were no other alternatives [53]. However, still today, they retain the best quality and completeness of the data along different dimensions [52], and for this reason, they are the most used for bibliometric analysis. The choice between the two is not neutral, and it conditions the final results [15]; differences between these databases are still not completely clear, and they are contested [54].

Some authors showed that Scopus has a general broader coverage than WoS [52], [55]. Despite its broader coverage, Scopus presents some issues. Firstly, Scopus does not provide pre-processed standardized reference lists as WoS does. Indeed, Scopus returns the complete references as written on the documents. It is a significant problem since the same reference is usually written in different ways in different documents. Thus, to use them for bibliographic analysis, they need a complex and error-prone matching procedure. Moreover, Scopus's higher coverage (relative to WoS) consists of low citation impact and more nationally oriented journals or conferences, leading to an over-representation of geographically-specific or less-influential literature [54].

For the aforementioned reasons, we collected the data for this work on the WoS database with the following query:
- **Topic**: ((precision OR smart OR digital OR intelligent) NEAR/1 (agriculture OR farming)) OR "agriculture 4.0"
- **Timespan**: 1985-2019
- **(Refined by) Languages**: English
- **Indexes**: SCI-EXPANDED, SSCI, A&HCI, CPCI-S, CPCI-SSH, BKCI-S, BKCI-SSH, ESCI, CCR-EXPANDED, IC

In this way, we retrieved 7257 documents on the 15th of December 2020.

The first part of the query contains terms related to the digitalization process in agriculture. These terms are connected with the NEAR/*n* operator to the second part of the query. The NEAR/*n* operator allows searching for terms that are separated by at most $n$ words. For example, with the operator NEAR/1, we retrieved results like "Precision Livestock Farming". "Precision" must be separated by "Farming" by no more than one word. The higher $n$ is, the more unrelated studies are retrieved. After few tests, we found that $n = 1$ was a good compromise between the total number of studies retrieved and the number of unrelated studies.

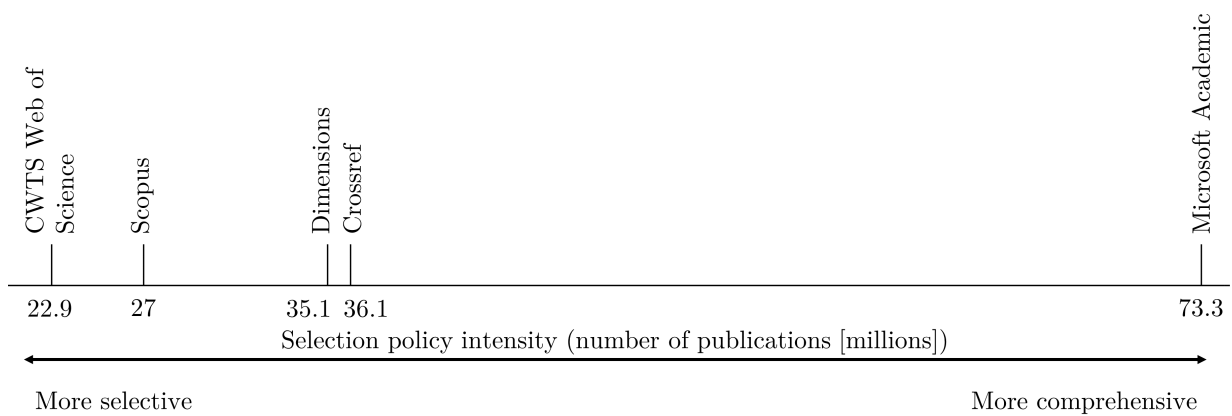

**FIGURE 2.** Selection policy intensity of the most common publications databases plotted on a line. The number of publications of each database is used as a proxy for the intensity. CTWS Web of Science is a subset of all the citation indices (SCIE, SSCI, AHCI, CPCI) to which CTWS research center had access (the plot has been adapted from [52]).

#### 3) DATA PREPARATION

In general, it is not possible to directly use the data collected from the publications databases. Most of the time, raw data need a pre-processing phase before being ready for the analysis. Using data without any cleaning or pre-processing phase could lead to worthless results. Indeed, the quality of the results depends on the quality of the data [15].

The first operation we did was to remove rows containing missing values in necessary variables. Web of Science data is already pre-processed and standardized in many fields. However, the authors' keywords needed further processing before use. Indeed, different authors use different keywords to express the same concept. To solve this problem, we first performed lemmatization the process of reducing a word to its base form); secondly, we matched the acronyms with their full forms; and lastly, we matched the synonyms.

Finally, we removed the terms used in the query, such as "Precision Agriculture", "Smart agriculture", "Agriculture 4.0", and all the others, since we wanted to map topics regardless of what terms the authors use to refer to the DAR.

#### 4) DATA ANALYSIS

Bibliometrics is an ensemble of many methods: citation analysis, co-citation analysis, bibliographic coupling, co-author analysis, co-word analysis. Each method has its pros and cons, the proper unit of analysis, and specific objectives [14]. Only some of the bibliometrics methods were exploited in this work, and they are briefly explained.

Some of the analyses are based on citation counts. Citations are a measure of a publication's influence. This influence can be associated with the publication elements such as source, authors' institutions, authors' countries. In this way, it is possible to find the key articles and the most influential sources, institutions, and countries. However, newer publications have less time to be cited. Therefore, citation counts are biased toward older publications.

When considering a subset of an entire publication database, citations can be calculated locally or globally. Local Citations (LC) are computed by only considering citing articles of the local subset. Global Citations (GC) are computed by considering the citing articles of the entire publications database (in our case, WoS). In bibliometric studies, LC are usually preferred since they come from articles of the same research field. Thus, they are considered more relevant than GC.

Other impact indicators are based on the number of publications. It is then possible to build rankings of the most productive sources, institutions, and countries. However, a higher number of publications does not necessarily imply a higher contribution to the research field's advancement. Thus, a commonly used citation impact indicator is the average number of citations per publication of a research unit. If the research unit is a journal, the indicator is called Impact Factor (IF).

In this study, we reported the Journal Citation Reports (JCR) IF computed by Clarivate Analytics. The JCR IF is a ratio between citations and recent citable items published. Thus, it is calculated by dividing the number of current year citations to the source items published in a journal during the previous two years by the number of publications in those two years. For example, the 2019 JCR IF of journal X is calculated by dividing the total number of citations from 2019 articles to 2017-2018 articles of journal X by the total number of publications in 2017-2018 of journal X.

Beyond the basic impact indicator, we also performed a collaboration and co-word analysis. Collaboration analysis is based on publications co-authorships. In our case, we aimed at mapping collaborations between countries. Thus, if a couple of a publication's authors belong to institutions in different countries, this is considered a collaboration between those two countries. Co-word analysis, instead, is based on the co-occurrence of terms in the same publication.

Both collaboration and co-word analyses are based on co-occurrence network analysis. A co-occurrence network is represented by a matrix:

$$B = A^T A,$$

where $B$ is the co-occurrence network matrix, and $A$ is a binary matrix representing a bipartite network *Documents* x *Attribute*. *Attribute* equals *Author* for a collaboration network and *Keyword* for a co-word network.

For collaboration networks, the generic element $a_{ij}$ of matrix $A$ is equal to 1 if document $i$ has been authored by author $j$, 0 otherwise. Instead, the generic element $b_{ij}$ of matrix $B$ indicates the number of collaborations between author $i$ and author $j$. The diagonal element $b_{ii}$ indicates the number of documents authored or co-authored by researcher $i$.

For co-word networks, the generic element $a_{ij}$ of matrix $A$ is equal to 1 if the document $i$ contains keyword $j$, 0 otherwise. Instead, the generic element $b_{ij}$ of matrix $B$ indicates the number of co-occurrences of keyword $i$ and keyword $j$. The diagonal element $b_{ii}$ is the number of documents in which the keyword $i$ appeared.

Once the network is built, it is clustered by Louvain algorithm [56]. Louvain clustering is a simple method to extract the community structure of large networks. It consists of a greedy optimization of a value called modularity. Modularity measures the density of links inside communities compared to links between communities. Once we have clustered the collaboration network, we can plot it. For co-word analysis, we plotted the network clusters on a thematic diagram [57]. Clusters are positioned on the diagram based on Callon's centrality and density measures [58]. Callon's centrality reads as the relative importance of a topic compared to the others. Callon's density reads as the development, in terms of numerousness of publications, of a topic. Each of the topics (or themes) is then classified based on its positioning in one of the four quadrants composing the thematic diagram:

- Upper-right quadrant - Motor themes: well-developed and important themes that structure a research field. They have a high density and centrality.
- Upper-left quadrant - Niche themes: themes that are well developed but are on the research field's borders. They present a high density but low centrality.
- Lower-left quadrant - Emerging or declining themes: themes characterized by low density and centrality.
- Lower-right quadrant - Basic themes: transversal themes in the research field but less developed than others. Basic themes are characterized by low density but high centrality.

5) DATA VISUALIZATION AND INTERPRETATION

Data visualization consists of graphically showing the results of the bibliometric analysis. Depending on the data, many kinds of visualization can be used: bar plots, line plots, networks, diagrams, and others. Once a proper visualization format is chosen, the results are interpreted by the researchers.

Prior knowledge of the research field is necessary for interpreting the findings. Researchers with in-depth knowledge have a clear advantage. However, they must not fit their preconceptions to the bibliometric analysis results, but the opposite. They should use their prior knowledge to enhance the findings [14].

## IV. DATA ANALYSIS AND INTERPRETATION

In this section, we show the results of the bibliometric analysis. We made a preliminary analysis of the entire dataset with the publications in the time span 1985-2019. Since we identified that DAR related studies began their diffusion in 2012, we then restricted the analysis to the time span 2012-2019. We first provide some general information about the data. Afterward, we list the most influential sources and institutions according to citation impact indicators. We also list the most cited articles and the most influential countries in terms of contributions and citation impact. To further understand the geographical dynamics, we analyzed the countries' collaboration network. Finally, we identified five main research streams, and we give a broad overview of them by analyzing the key articles for each. We conclude by presenting the trending topics of the last three years (2017, 2018, and 2019).

### A. PRELIMINARY ANALYSIS

Figure 3 shows the year-wise distribution of the number of publications in the period 1993-2019 (1993 is the year of the first publication in the dataset). The annual growth rate was on average 30.5%. We can spot a turning point in 2014 when scientific production started to increase faster. The application of digital technologies in agriculture likely opened new research possibilities. In 2019, the DAR annual scientific production reached the number of 1324 documents. Such a large amount of studies requires the use of algorithmic bibliometric tools to be analyzed.

To restrict the analysis to a proper time span, we made a plot of the year-wise frequency of the ten most frequent authors' keywords in Figure 4. In the figure, it is possible to recognize three groups of terms. A group of terms began their diffusion in the '90s; it then reached a plateau, if not a decline: Spatial Variability, Geostatistic, SSM, Variable Rate Application, GPS, and GIS. These terms are mostly related to PA.

Another group of terms began their diffusion around 2012: IoT, CSA, Machine Learning (ML), and UAS (that began its diffusion in 2010). Since these technologies and frameworks are pillars of the DAR, we decided to take the year 2012 as the beginning of our time span (2012-2019). Finally, the third group of terms began its diffusion before 2010 and saw new popularity since 2012: RS, Hyperspectral Imaging, Vegetation Index, and Sensor. The renewed diffusion of these terms is likely due to the introduction of UAS in agriculture and on-the-go proximal sensing systems.

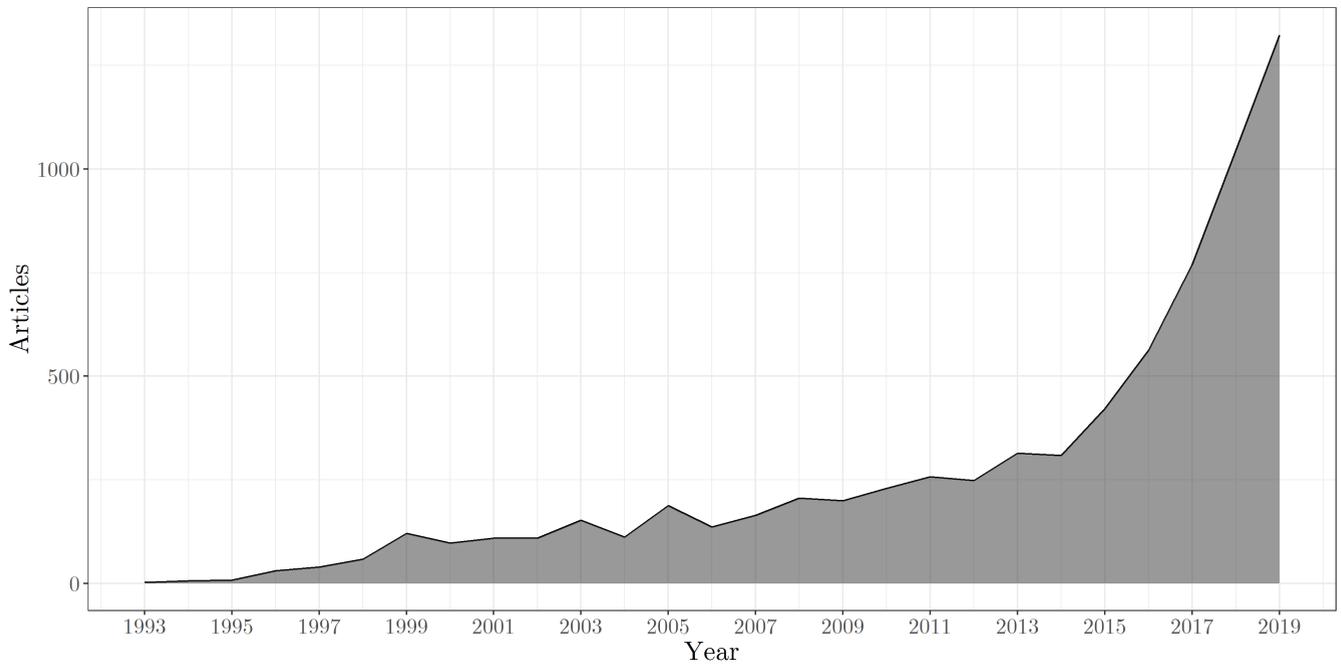

**FIGURE 3.** DAR annual scientific production in the period 1993-2019.

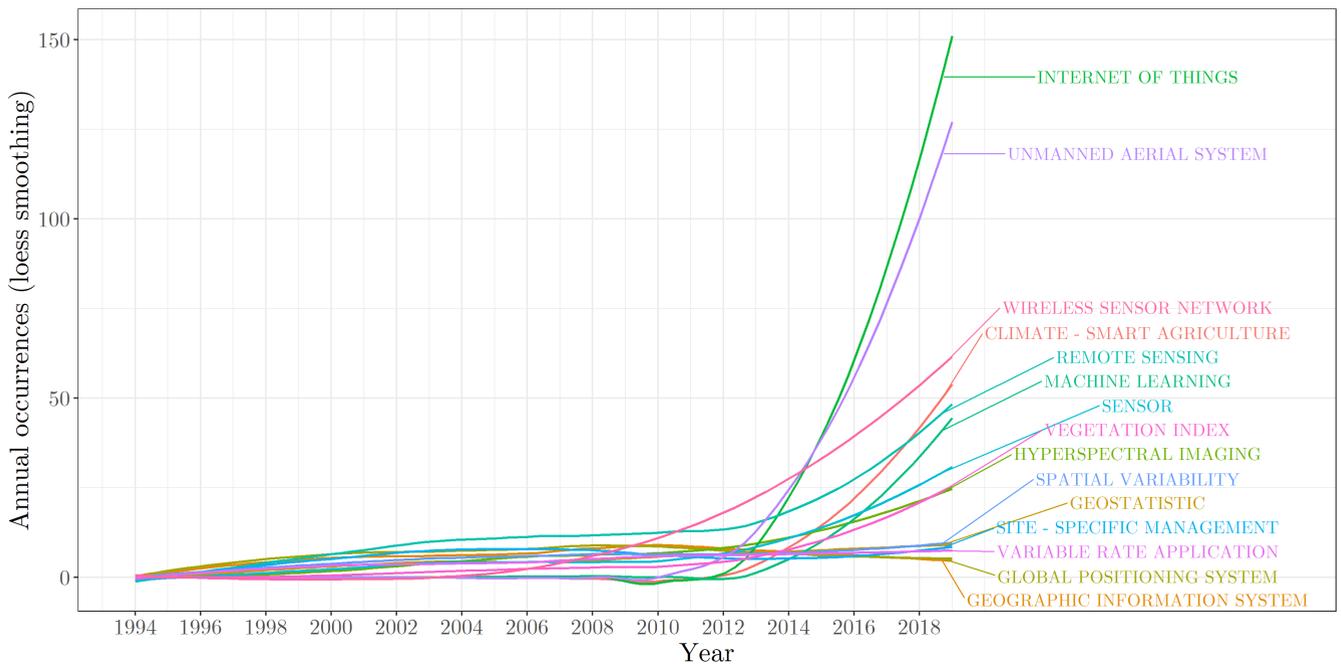

**FIGURE 4.** Top ten authors' keywords growth. The graph shows the smoothed annual occurrences of the ten most frequent authors' keywords used in the DAR publications.

### B. GENERAL INFORMATION

In Table 1 we give some general information about the data. The subset 2012-2019 contains 4995 documents from 2152 sources. The publications derive from 15139 authors, and the average number of authors per document is 3.03. The average citations per document are 10.03, and the total number of unique authors' keywords is 10333.

The percentage composition by document types is illustrated in Figure 5. The documents are for most articles (53.59%). It has to be noted the high percentage (36.38%) of proceedings papers that are characteristic of some research areas such as computer science. Reviews count for 5.16% and book chapters for 2.83%.

**TABLE 1.** General information about the dataset we used in our analysis.

| General information about the data | |
|---|---|
| Timespan | 2012:2019 |
| Sources (Journals, Books, etc.) | 2152 |
| Documents | 4995 |
| Authors | 15139 |
| Authors per Document | 3.03 |
| Average citations per document | 10.03 |
| Average citations per year per doc | 1.81 |
| References | 124277 |
| Authors' Keywords | 10333 |

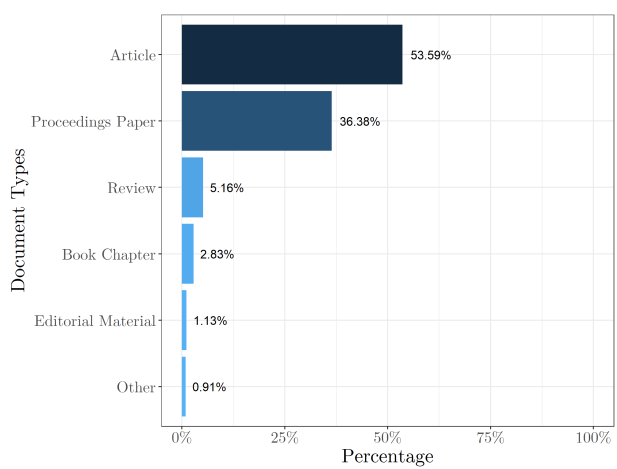

**FIGURE 5.** Percentage composition by document types of DAR related publications.

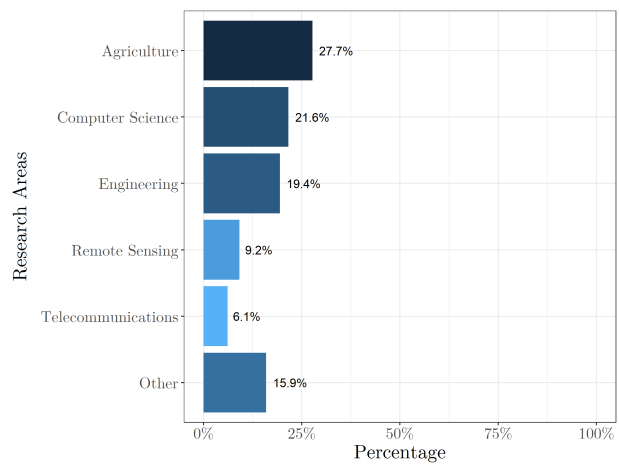

**FIGURE 6.** Percentage composition by research area of DAR related publications.

Figure 6 shows the percentage composition by research area. Of course, the principal research area is agriculture, with 27.7% on the total number of documents. The other main research areas are computer science (21.6%), engineering (19.4%), remote sensing (9.2%), and telecommunications (6.1%). It is possible to note a clear prevalence of research areas related to digital technologies.

### C. MOST INFLUENTIAL SOURCES, INSTITUTIONS, ARTICLES AND COUNTRIES

Table 2 shows the most influential sources (journals, conferences, books, etc.). We reported the most relevant sources by the number of documents ranked according to the Standard Competition Ranking (SCR). In SCR, items that compare equal receive the same ranking number, and then a gap is left in the ranking numbers. The first item that does not compare equal to a sequence of two or more equally ranked items receives a rank equal to its ordinal position. We also reported the 2019 JCR IF for journals listed in the same table.

The top three relevant sources by number of documents are "Computers and Electronics in Agriculture" (259 documents - 5.11%), "Precision Agriculture" (143 documents - 2.82%), and "Sensors" (132 documents - 2.61%). Among the top twenty relevant sources, the top three 2019 JCR IF sources are "Geoderma" (4.848), "Remote Sensing" (4.509), and "Precision Agriculture" (4.454). In Table 2 we also made a SCR of the most local cited journals. The top three ranked journals are "Computers and Electronics in Agriculture" (6290), "Remote Sensing of Environment" (4318), and "Precision Agriculture" (3739).

Table 3 shows the most productive institutions. We made a SCR by ordering institutions according to the number of published documents. The number of published documents per institution has been computed with the Organizations-Enhanced WoS field. Indeed, institutions usually have many name variants. If these variations are not considered, the total counts do not reflect an institution's real productivity. The Organizations-Enhanced WoS data are computed by merging the different variations of an institution's name. The number of documents of systems of institutions contains the number of documents of single institutions. For example, the 60 documents from the University of Florida are contained in the 73 documents from the State University System of Florida. The top five productive institutions are the "United States Department of Agriculture USDA" (133 documents), the "CGIAR" (120 documents), the "Chinese Academy of Sciences" (103 documents), the "Wageningen University Research" (93 documents), and the "Consejo Superior de Investigaciones Cientificas CSIC" (88 documents). CGIAR is a global partnership of many international organizations, like the FAO and the World Bank, engaged in food security research.

Table 4 shows the ten most local cited articles. The first ranked article is "David J. Mulla" [6] with 180 LC, followed by "Wolfert et al." [9] with 93 LC and "Bendig et al." [59] with 58 LC. These three articles are also the most globally cited. We discuss in detail some of the articles listed in Table 4 in the remainder of this section.

The twenty most productive countries are shown in Figure 7. The country associated with each publication is the institution's country of the corresponding author.

**TABLE 2.** Most relevant sources ordered according to the Standard Competition Ranking. Each relevant source has associated its 2019 JCR IF. Sources for which a JCR IF was not available has been given the "NA" value. The most cited sources has been ordered by LC.

| SCR by N. of documents | Relevant Sources | N. of documents | % | 2019 JCR IF | SCR by LC | Most Cited | LC |
|---|---|---|---|---|---|---|---|
| 1 | COMPUTERS AND ELECTRONICS IN AGRICULTURE | 259 | 5.11% | 3.858 | 1 | COMPUTERS AND ELECTRONICS IN AGRICULTURE | 6290 |
| 2 | PRECISION AGRICULTURE | 143 | 2.82% | 4.454 | 2 | REMOTE SENSING OF ENVIRONMENT | 4318 |
| 3 | SENSORS | 132 | 2.61% | 3.275 | 3 | PRECISION AGRICULTURE | 3739 |
| 4 | REMOTE SENSING | 122 | 2.41% | 4.509 | 4 | AGRONOMY JOURNAL | 2339 |
| 5 | PROCEEDINGS OF SPIE | 113 | 2.23% | NA | 5 | REMOTE SENSING | 2119 |
| 6 | BIOSYSTEMS ENGINEERING | 60 | 1.18% | 3.215 | 6 | BIOSYSTEM ENGINEERING | 1950 |
| 7 | IEEE INTERNATIONAL SYMPOSIUM ON GEOSCIENCE AND REMOTE SENSING IGARSS | 55 | 1.09% | NA | 7 | SENSORS | 1802 |
| 8 | INTERNATIONAL CONFERENCE ON AGRO GEOINFORMATICS | 51 | 1.01% | NA | 8 | GEODERMA | 1664 |
| 9 | GEODERMA | 41 | 0.81% | 4.848 | 9 | TRANSACTIONS OF THE ASABE | 1514 |
| 10 | SUSTAINABILITY | 40 | 0.79% | 2.576 | 10 | INTERNATIONAL JOURNAL OF REMOTE SENSING | 1500 |
| 11 | TRANSACTIONS OF THE ASABE | 38 | 0.75% | 1.156 | 11 | SOIL SCIENCE SOCIETY OF AMERICA JOURNAL | 1475 |
| 12 | INTERNATIONAL ARCHIVES OF THE PHOTOGRAMMETRY REMOTE SENSING AND SPATIAL INFORMATION SCIENCES | 37 | 0.73% | NA | 12 | FIELD CROPS RESEARCH | 1377 |
| 13 | ENGENHARIA AGRICOLA | 36 | 0.71% | 0.603 | 13 | JOURNAL OF DAIRY SCIENCE | 1164 |
| 14 | APPLIED ENGINEERING IN AGRICULTURE | 34 | 0.67% | 0.973 | 14 | AGRICULTURAL SYSTEMS | 1121 |
| 14 | IFAC PAPERSONLINE | 34 | 0.67% | NA | 15 | SOIL & TILLAGE RESEARCH | 1052 |
| 14 | INTERNATIONAL JOURNAL OF AGRICULTURAL AND BIOLOGICAL ENGINEERING | 34 | 0.67% | 1.731 | 16 | AGRICULTURAL WATER MANAGEMENT | 989 |
| 17 | AGRICULTURAL SYSTEMS | 33 | 0.65% | 4.212 | 17 | IEEE TRANSACTIONS ON GEOSCIENCE AND REMOTE SENSING | 968 |
| 17 | IEEE ACCESS | 33 | 0.65% | 3.745 | 18 | AGRICULTURE, ECOSYSTEMS & ENVIRONMENT | 888 |
| 19 | AGRONOMY BASEL | 31 | 0.61% | 2.603 | 19 | SCIENCE | 877 |
| 19 | LECTURE NOTES IN COMPUTER SCIENCE | 31 | 0.61% | NA | 20 | PLOS ONE | 846 |

**TABLE 3.** Most productive institutions in terms of number of publications. The number of documents of systems of institutions contains the number of documents of single institutions.

| SCR by N. documents | Organization | N. of documents | SCR by N. of documents | Organization | N. of documents |
|---|---|---|---|---|---|
| 1 | UNITED STATES DEPARTMENT OF AGRICULTURE USDA | 133 | 11 | UNIVERSIDADE DE SAO PAULO | 57 |
| 2 | CGIAR | 120 | 12 | ALLIANCE | 51 |
| 3 | CHINESE ACADEMY OF SCIENCES | 103 | 13 | BEIJING ACADEMY OF AGRICULTURE FORESTRY | 50 |
| 4 | WAGENINGEN UNIVERSITY RESEARCH | 93 | 14 | UNIVERSITY OF BONN | 49 |
| 5 | CONSEJO SUPERIOR DE INVESTIGACIONES CIENTIFICAS CSIC | 88 | 15 | INDIAN COUNCIL OF AGRICULTURAL RESEARCH ICAR | 48 |
| 6 | CHINA AGRICULTURAL UNIVERSITY | 86 | 16 | COMMONWEALTH SCIENTIFIC INDUSTRIAL RESEARCH ORGANISATION CSIRO | 47 |
| 7 | STATE UNIVERSITY SYSTEM OF FLORIDA | 73 | 16 | INRAE | 47 |
| 8 | CONSIGLIO NAZIONALE DELLE RICERCHE CNR | 65 | 18 | INTERNATIONAL CENTER FOR TROPICAL AGRICULTURE CIAT | 44 |
| 9 | UNIVERSITY OF FLORIDA | 60 | 19 | EMPRESA BRASILEIRA DE PESQUISA AGROPECUARIA EMBRAPA | 43 |
| 10 | UNIVERSITY OF CALIFORNIA SYSTEM | 58 | 19 | UNIVERSITY OF NEBRASKA LINCOLN | 43 |

A country's total number of publications is divided into Single Country Publications (SCP) and Multiple Country Publications (MCP). Thus, if a publication's authors are all of the same nationality as the corresponding author, this publication shows as an SCP for that nation; vice versa, it shows as an MCP. The most productive countries are the USA, with 653 publications (of which 524 SCP and 129 MCP), followed by China with 621 publications (of which 489 SCP and 132 MCP), and India with 445 publications (of which 403 SCP and 42 MCP). Among these twenty countries, the most internationally collaborating country is the United Kingdom with an MCP ratio (MCP/TP, where TP is the total number of publications) of 0.47, followed by Belgium with an MCP ratio of 0.41 and Canada with an MCP ratio of 0.36. It has to be noted that the number of a country's publications can be biased towards its population size.

Figure 8 shows the twenty highest cited countries. The first-ranked country is the USA with 7351 total citations, followed by Spain with 4821 citations and China with 4402. Among these twenty countries, the top three for average article citations are the Netherlands, with 26.6 citations per article, Finland with 22.5 citations, and Spain with 21.2 citations.

**TABLE 4.** Most cited articles according to LC counts. LC are those received from articles in our dataset. GC are those received from articles in the entire WoS database.

| SCR by LC | Authors(s) | Year | Title | LC | GC |
|---|---|---|---|---|---|
| 1 | David J. Mulla | 2013 | Twenty five years of remote sensing in precision agriculture: Key advances and remaining knowledge gaps | 180 | 553 |
| 2 | Sjaak Wolfert, Lan Ge, Cor Verdouw, Marc-Jeroen Bogaardt | 2017 | Big Data in Smart Farming – A review | 93 | 350 |
| 3 | Juliane Bendig, Andreas Bolten, Simon Bennertz, Janis Broscheit, Silas Eichfuss, Georg Bareth | 2014 | Estimating Biomass of Barley Using Crop Surface Models (CSMs) Derived from UAV-Based RGB Imaging | 58 | 239 |
| 4 | Eija Honkavaara, Heikki Saari, Jere Kaivosoja, Ilkka Pölönen, Teemu Hakala, Paula Litkey, Jussi Mäkynen, Liisa Pesonen | 2013 | Processing and Assessment of Spectrometric, Stereoscopic Imagery Collected Using a Lightweight UAV Spectral Camera for Precision Agriculture | 55 | 224 |
| 5 | José Manuel Peña, Jorge Torres-Sánchez, Ana Isabel de Castro, Maggi Kelly, Francisca López-Granados | 2013 | Weed Mapping in Early-Season Maize Fields Using Object-Based Analysis of Unmanned Aerial Vehicle (UAV) Images | 52 | 163 |
| 6 | Jorge Torres-Sánchez, José Manuel Peña, Ana Isabel de Castro, Francisca López-Granados | 2014 | Multi-temporal mapping of the vegetation fraction in early-season wheat fields using images from UAV | 50 | 173 |
| 7 | Sebastian Candiago, Fabio Remondino, Michaela De Giglio, Marco Dubbini, Mario Gattelli | 2015 | Evaluating Multispectral Images and Vegetation Indices for Precision Farming Applications from UAV Images | 49 | 202 |
| 8 | Mare Srbinovska, Cvetan Gavrovski, Vladimir Dimcev, Aleksandra Krkoleva, Vesna Borozan | 2015 | Environmental parameters monitoring in precision agriculture using wireless sensor networks | 48 | 179 |
| 9 | Anne-Katrin Mahlein | 2016 | Plant Disease Detection by Imaging Sensors - Parallels and Specific Demands for Precision Agriculture and Plant Phenotyping | 47 | 238 |
| 10 | Jakob Geipel, Johanna Link, Wilhelm Claupein | 2014 | Combined Spectral and Spatial Modeling of Corn Yield Based on Aerial Images and Crop Surface Models Acquired with an Unmanned Aircraft System | 43 | 119 |

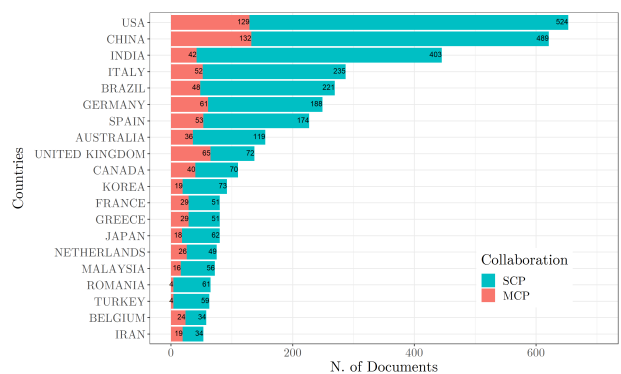

**FIGURE 7.** Most productive countries in the DAR research field in terms of number of publications. Each country's productivity is divided in two parts, Single Country Publications (SCP) and Multiple Country Publications (MCP).

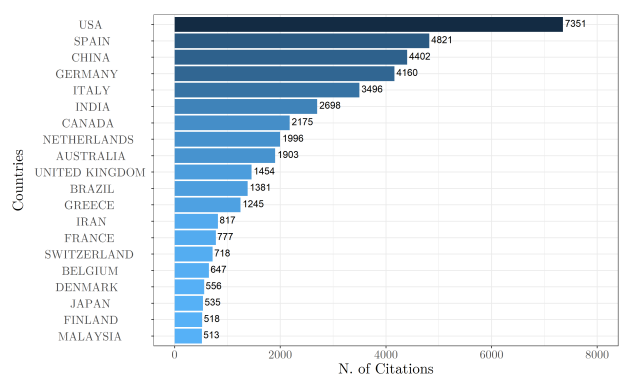

**FIGURE 8.** Most cited countries in the DAR research field. The number of citations is the sum of all the GC received by publications of a certain country.

Figure 9 shows a clustered network of countries collaborations in a circular shape. The size of edges is proportional to the number of collaborations between countries. A collaboration between two countries comes from the co-authorship of two different authors' countries. The size of the network's nodes is proportional to the number of publications of that country. Also, we imposed the number of countries to twenty.

From the network of Figure 9 we can recognize two main clusters: a European cluster dominated by Italy, Germany, Spain, and the United Kingdom; a ''multi-continent'' cluster dominated by the USA, China, India, and Brazil. We also recognize a third minor cluster composed of Iran and Malaysia with 3 collaborations between them. Likely, the last cluster will increase in size by imposing a higher number of countries in the network. The most robust collaboration worldwide is between the USA and China with 99 collaborations, followed by the USA and Canada with 39 collaborations and the USA and the United Kingdom with 32 collaborations.

In the European cluster, the most robust collaborations are between Italy and Germany with 21 collaborations, Italy and Netherlands with 21 collaborations, followed by Italy and United Kingdom with 19 collaborations. In the ''multi-continent'' cluster, the most robust collaboration is between the USA and China (99 collaborations), the USA and Canada (39 collaborations), and the USA and India with 31 collaborations.

### D. MAIN RESEARCH STREAMS

In this section, we elaborate on the thematic diagram showed in Figure 10. The thematic diagram has been built (as explained in Section III-B4) from the clustering of a co-word network. Each node of the network is an authors' keyword, and the degree of an edge is proportional to the number of co-occurrences of two words (vertices) in publications. Then, the network has been clustered via the Louvain algorithm, and we obtained five clusters: CSA, SSM, RS, IoT, and AI. We have attributed the clusters' names based on the clusters-related terms. The ten most frequent words associated with each cluster are reported in Table 5.

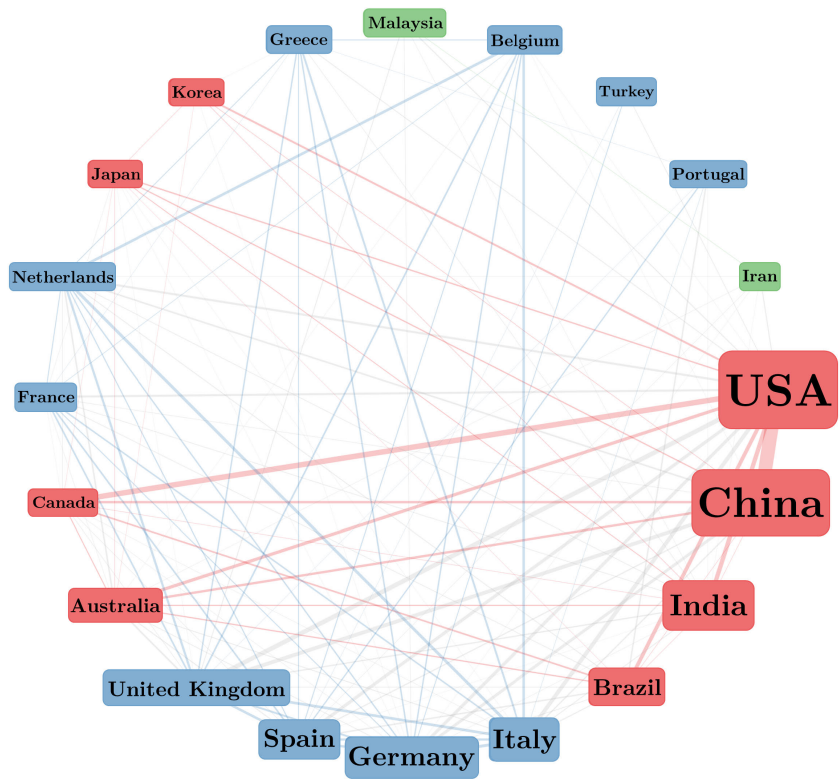

**FIGURE 9.** Countries collaboration network. Nodes size is proportional to the number of a country's publications. Edge size is proportional to the number of collaborations between two countries. Nodes are clustered via Louvain algorithm.

In the following, we give a broad overview of each cluster by presenting the most relevant publications. The most relevant publications are those with the highest Cluster Relevance (*CR*) score. The *CR* of publication *i* for cluster *j* is:

$$CR_{ij} = LC_i * n\_matches_{ij},$$

where $LC_i$ is the number of local citations of publication *i*, and $n\_matches_{ij}$ is the number of common authors' keywords between publication *i* and cluster *j* related keywords.

We analyzed at least five publications between those with the highest *CR* preferring reviews as document types for each cluster. We analyzed more publications if we deemed it was necessary to cover some key aspects that remained untreated by the five publications selected at first. In the following, we first present the concept of CSA since its objectives are shared by the majority of the DAR studies. Then, we present the SSM cluster since it contains literature that has been historically developed first. After, we present the RS and IoT clusters as they are the two motor themes. Finally, we conclude by presenting the AI cluster that has been categorized as an emerging theme on the thematic diagram.

1) CLIMATE-SMART AGRICULTURE

CSA is the set of agricultural practices to adapt to climate changes, reduce GHG emissions, and promote a sustainable intensification for food security [60]. These three pillars of CSA are usually condensed in the terms adaptation (or resilience), mitigation, and productivity (sustainable intensification) [61].

The authors of [61] made a systematic literature review on the institutional perspectives of CSA. They found that mitigation of GHG was predominantly addressed in high-income countries, while adaptation and productivity were prioritized in middle and low-income countries. Similarly, in [5], they found that mitigation options were less likely profitable for smallholder farmers.

CSA has roots in the 2007 Intergovernmental Panel on Climate Change (IPCC). Indeed, the IPCC (2007) clearly stated the dual nature of the agricultural sector, that is, being at the same time a significant contributor to global GHG emissions and subject to the threats of climate changes. It became then evident the need for a transition to climate-friendly agricultural practices. CSA was then formalized in 2010 by the FAO and the World Bank [5].

CSA has received some critiques that remain unfolded [62]. The main critique of CSA is the lack of clear attributes and performance criteria for an agricultural practice to be considered part of CSA. Without such clarity, it is easy for unsustainable practices to be labeled as CSA. In the literature, this risk is referenced with the term green-washing [5].

Another concern is that scientists' attention is centralized on technological advancement while social, institutional, and economic issues are poorly considered [5]. In [62] and [61] the authors highlighted that a technology-push approach does not always lead to the expected outcomes because of underestimating the local context. Thus, they underscored the importance of a paradigm shift from a technology-oriented approach to a system-oriented approach.

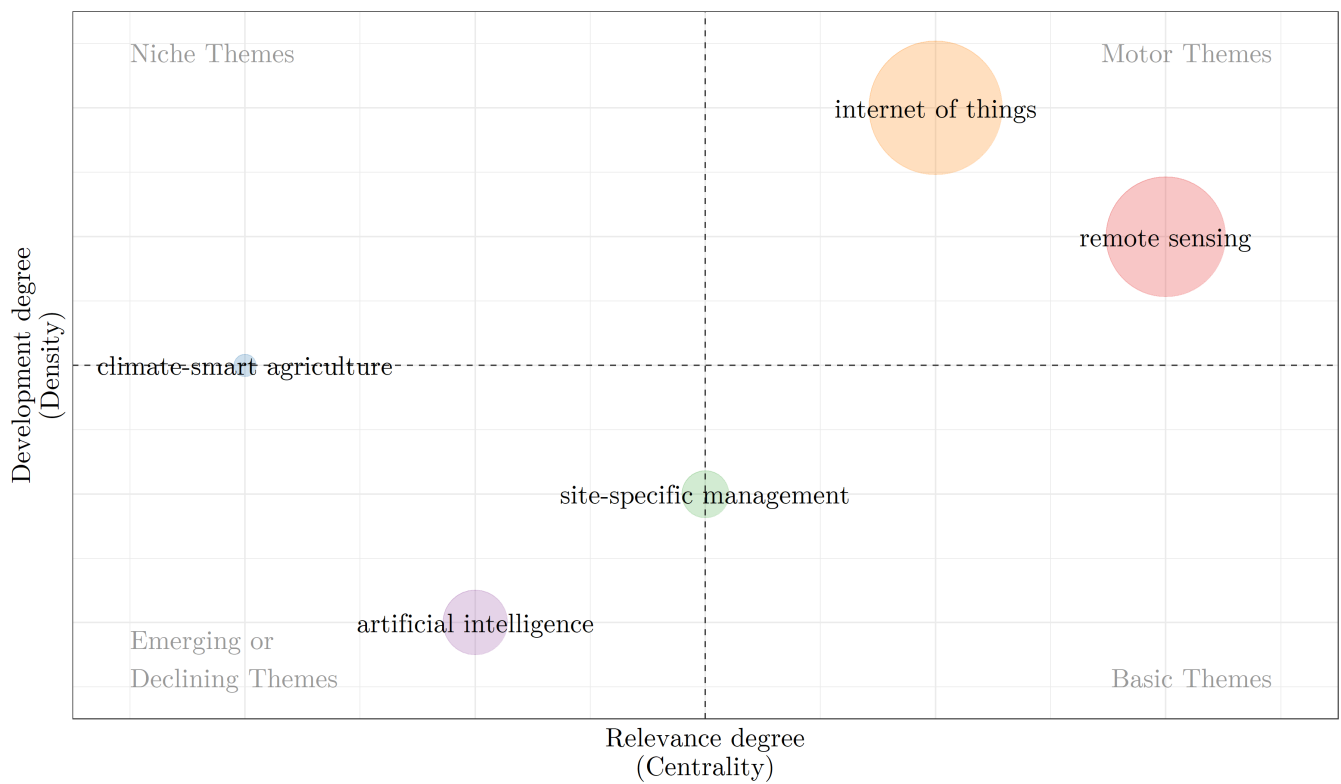

**FIGURE 10.** Thematic diagram of the DAR research field. Each circle represents a cluster and its size is proportional to the sum of all cluster's words occurrences. Clusters are placed on the diagram based on their Callon's centrality and density measures.

**TABLE 5.** List of the top ten keywords by the number of occurrences for each of the thematic clusters.

| 1 - Climate-Smart Agriculture | Occurences | 2 - Site-Specific Management | Occurences | 3 - Remote Sensing | Occurences | 4 - Internet of Things | Occurences | 5 - Artificial Intelligence | Occurences |
|---|---|---|---|---|---|---|---|---|---|
| climate - smart agriculture | 161 | geostatistic | 65 | unmanned aerial system | 367 | internet of things | 433 | machine learning | 117 |
| climate change | 88 | spatial variability | 59 | remote sensing | 205 | wireless sensor network | 293 | precision livestock farming | 101 |
| food security | 44 | variable rate application | 52 | hyperspectral imaging | 117 | sensor | 133 | image processing | 83 |
| adaptation | 39 | geographic information system | 50 | vegetation index | 100 | big data | 66 | deep learning | 70 |
| sustainability | 30 | global positioning system | 46 | normalized difference vegetation index | 80 | cloud computing | 59 | computer vision | 59 |
| mitigation | 27 | management zone | 44 | multispectral imaging | 62 | decision support system | 57 | artificial neural network | 50 |
| sustainable agriculture | 26 | site - specific management | 44 | leaf area index | 46 | zigbee | 42 | convolutional neural network | 38 |
| adoption | 23 | yield mapping | 42 | plant phenotyping | 37 | irrigation | 40 | classification | 37 |
| optimization | 18 | global navigation satellite system | 33 | nitrogen | 33 | automation | 39 | machine vision | 37 |
| integrated pest management | 17 | soil | 31 | partial least squares regression | 26 | soil moisture | 38 | support vector machine | 31 |

Also, the DAR could further increase the digital divide between the different countries and social groups. If not properly regulated, digital technologies can translate into the consolidation of power of few companies that own the data [29]. A literature gap consists in the lack of research about the synergies and trade-offs among the three CSA pillars (productivity, adoption, and mitigation) [61], [63].

A branch of the CSA research field is dedicated to studying the barriers and drivers of the adoption of DAR technologies. In particular, a focus is made on the economic, sociological, environmental, and entrepreneurial aspects. The most important factors influencing the adoption of DAR technologies are explained in [64].

#### 2) SITE-SPECIFIC MANAGEMENT

SSM is a topic that lies between declining and basic themes in the thematic map of Figure 10. It is interesting to note that this cluster's studies use the term PA rather than Smart or Digital Agriculture. The SSM literature is focused on spatial variability with particular attention on soil variability. Typical data analysis tools are those from the field of geostatistic. SSM techniques aim to define crop field sub-units to manage them in a targeted way. Then, field sub-units (or Homogeneous Management Zones - HMZ) are handled with Variable Rate Applications of substances (water, fertilizers, pesticides). Popular SSM technologies are the GNSS, the GIS, and yield mapping.

SSM sees its roots in the birth of PA during the mid-1980s. At that time, two main practices were diffusing. One of these was "farming by soil", which consisted of estimating the field's spatial variability by analyzing samples from soil units. At the same time, a practice that we now call SSM was developing. In SSM, fields are divided into homogeneous sub-units to receive customized treatments. In the years, SSM superseded the other practice since it showed the advantages of a finer sub-units mapping compared to the larger and more heterogeneous soil units of the "farming by soil" practice [6].

SSM was at first conducted with RS technologies, like satellite imaging. The aim was to map the organic matter, soil phosphorus, and crop yield for defining HMZs. More recently, sensors are mounted on tractors, UGV, or UAV to detect soil properties. A breakthrough was brought by apparent electrical conductivity ($EC_a$) sensors. Apparent electrical conductivity is a parameter influenced by many soil physical-chemical properties [65]. In [66], they found principal-components groups (consisting of soil properties and some exchange cations) that explained >50% of the variability in $EC_a$.

One of the few SSM studies performed on large-scale data is that of [67]. The authors analyzed multiple-years data of 571 fields, from 110 farmers, in eight different USA states and regarding four different crops (corn, soybean, wheat, and cotton). They aimed at analyzing spatial patterns by correlating yield and four covariates (red band spectral reflectance, NDVI, plant surface temperature, and historical yields). In [68] they obtained higher net income and yields by reducing the environmental impact in a nitrate vulnerable zone. They divided a field into three HMZs by merging multiple data layers with geostatistical algorithms. They also exploited sand content, Soil Organic Matter, $EC_a$, and crop yield maps. The three HMZ were, in turn, divided into two sub-zones to compare uniform and variable-rate applications.

In [69], they reviewed the methods for improving Nitrogen Use Efficiency (NUE). NUE measures plants' efficiency in uptaking nitrogen compared to the amount left in the soil or lost. Worldwide, NUE is relatively low; world cereal crops have an average NUE of 33%. Reasons for low NUE are multiple and are explained in [69].

### 3) REMOTE SENSING

The RS topic has been categorized as a motor theme in the thematic map of Figure 10. RS consists of non-contact measurements of the reflected or emitted radiation from agricultural elements like plants or soil. Plants can emit energy in the form of fluorescence or thermal emission beyond reflected radiation [6]. The amount of reflected radiation is inversely related to the radiation absorbed by plant pigments like chlorophyll. Consequently, the amount of reflected radiation in specific wavelengths can be used to compute Vegetation Indices [6]. Vegetation Indices (also called Spectral Indices) are dozens; a comprehensive list and comparison can be found in [70].

Vegetation Indices are utilized for assessing plants' morphometric and physiological parameters. The measuring of these plant characteristics is called Plant Phenotyping. A complete review on sensors for Plant Phenotyping and measurable plant characteristics is that of [71].

There are different kinds of platforms for carrying RS sensors, such as satellites, UAV, tractors, UGV, and hand-held sensors. Measurements performed with ground platforms like tractors, UGV, or hand-held sensors are also known as Proximal Sensing [6].

The RS platforms and their associated imaging systems can be differentiated by the platform's altitude, spatial resolution, spectral resolution, and temporal resolution (or minimum return frequency). Satellite imaging is among the first technologies adopted in RS. Many satellites have been launched since the early 1970s; a list updated till 2013 can be found in [6]. The spatial and spectral resolutions of satellite imaging are of great relevance and depend on the specific application as explained in [72].

In recent years we noted a rising trend of works exploiting Sentinel-2 data. Sentinel-2 are two satellites (Sentinel-2A and Sentinel-2B) launched by the European Space Agency for earth observation. These satellites are equipped with multispectral sensors, including 13 spectral bands, with a spatial resolution ranging from 10 m to 60 m. In [73] they investigated the differences between uniform and variable nitrogen fertilization treatments in wheat by computing NDVI images from two Sentinel-2 spectral bands.

Despite the ease of retrieving data through satellite imaging, there are also some cons. Indeed, satellite imaging is limited by cloud cover and is most reliable when irradiance is relatively consistent. Other challenges are the correction of atmospheric interference and off-nadir view angles, the geo-rectification of pixels, and the calibration of raw digital numbers to true surface reflectance [6].

Given the limitations of satellite imaging, more emphasis has recently been given to UAS. UAS are also called, or include, UAV, Remotely Piloted Vehicles (RPV), Remotely Operated Aircrafts (ROA), or Remotely Controlled Helicopters (RC-Helicopters). UAS have higher flexibility than satellites since they allow higher spatial and temporal resolutions. UAS can reach up to a centimeters accuracy and near real-time acquisition at relatively low operational costs [74].

Limitations of UAS are the platform reliability, the limited sensor payload, and the image processing. Unstable vehicle positioning makes image processing difficult because of different spatial resolutions and different viewing angles. Also, the low flight altitude causes geometric distortion. Finally, aviation regulations are an obstacle that hinders the application of UAS.

Alternatives to UAS are ground vehicles such as UGV or tractors equipped with sensors. They have the same advantages of the UAS of high spatial resolution and real-time sensing. Moreover, they can carry a heavy payload sensor, and they are not affected by winds and turbulence. However, they are affected by unstable movements and vibrations due to the coarse ground of agricultural fields.

To date, the dominant imaging technologies for RS are RGB imaging, hyperspectral imaging, and multispectral imaging. Besides, Synthetic Aperture Radar (SAR) has been demonstrated to be useful for crop conditions assessment. The limited diffusion of SAR imaging is probably due to the costs, timing, and interpretation of the data [74]. Proximal Sensing also includes other sensing technologies such as LiDAR sensors, range cameras, fluorescence sensors, and thermography. For a thorough explanation, refer to [71].

In recent years, spectral sensing is gaining new attention thanks to the growing popularity of UAS systems. Spectral technology has improved, resulting in smaller and lighter sensors that can be mounted on UAV platforms. The main difference between hyperspectral and multispectral technologies is the number of bands and their width [75]. While multispectral imaging usually ranges between five to twelve bands (10-40 nm each), hyperspectral imaging consists of hundreds of bands arranged in narrowed bandwidth (1-10 nm each).

Spectral sensing is used for a variety of applications, including food quality and crop conditions assessment. However, multispectral data does not enable the reaching of the same level of detail as hyperspectral sensing. Thus, some details might pass unnoticeable by multispectral sensors. Along with this improvement, hyperspectral sensing also increases the data processing complexity. Hyperspectral data can be challenging to analyze in real-time with reduced computational resources. A review on hyperspectral data handling pipelines can be found in [75].

### 4) INTERNET OF THINGS

The IoT, together with RS, is another motor theme in the thematic diagram of Figure 10. The IoT has been defined by the International Telecommunication Union (ITU) -a specialized agency of the United Nations- as ''a global infrastructure for the information society enabling advanced services by interconnecting (physical and virtual) things based on, existing and evolving, interoperable information and communication technologies'' [76].

A more extended definition is that of Kranenburg [77] who defined the IoT as ''a dynamic global network infrastructure with self-configuring capabilities based on standard and interoperable communication protocols where physical and virtual ''things'' have identities, physical attributes, and virtual personalities and use intelligent interfaces, and are seamlessly integrated into the information network, often communicate data associate with users and their environments''.

The author of [25] made an in-depth and broad review on the IoT technologies for agriculture. This review article explains the many functional blocks composing IoT systems and the most common hardware platforms and wireless communication standards. Also, various case studies are presented.

In [10] the authors developed an IoT system and a related data analysis pipeline for improving the yield of lime and homegrown vegetables. They equipped three study fields with humidity and soil moisture sensors, and they get the temperature from a web service.

The typical agricultural IoT systems centralize the whole management in a single controller or computer (at the farmer's premises or elsewhere). However, these systems lack modularity, delocalization, and the flexibility provided by virtualization. A trending paradigm that is still not widely adopted in the agricultural IoT is fog and edge computing. This paradigm consists of improving the system reaction by moving the computation close to the end devices. The authors of [24] proposed a three-tier architecture composed of a local, an edge, and a cloud plane.

IoT sensors communicate most of the time through wireless technologies. Multiple interconnected IoT sensors are called Wireless Sensor Network (WSN). WSN nodes are battery-powered, and the energy consumption is a major concern in the WSN research field. The state-of-the-art energy-efficient schemes for agricultural applications are presented in [78].

The massive amounts of data produced by IoT devices and other sensing technologies are usually referred to with the term Big Data. The authors of [9] made a review of the state-of-the-art Big Data applications in agriculture. Also, they identified the socio-economic challenges to be addressed related to the use of Big Data.

IoT data are usually collected by a master node and conveyed to a centralized database. However, single and centralized databases could be more prone to cyber-attacks, asynchronous inaccurate data, censorship, data distortion, and scientific misconduct. An alternative way of storing data is through blockchain technology. A blockchain is organized as a linear sequence of small datasets called ''blocks'', containing timestamped batches of transactions. The transaction records are distributed across a network of computers or databases. The blockchain infrastructure is immutable and decentralized for transparent data management. However, scalability remains the fundamental problem within blockchain networks. There exist a trade-off between block processing time and network propagation [79].

Agricultural Big Data comes from many sources and requires ad-hoc analytics and management software called Farm Management Information System (FMIS). FMISs allow the integration of spatial and temporal historical data, real-time farm data, knowledge sources, and economic models into a coherent management information system [80]. [81] defined a FMIS ''as a planned system for the collecting, processing, storing and disseminating of data in the form of information needed to carry out the operations functions of the farm''.

In [80] the authors analyzed 141 commercial software packages, and they recognized eleven functions. The authors exploited these eleven functions to group the software packages into four clusters. Then they compared the main characteristics of each cluster and highlighted the lacks and weaknesses.

### 5) ARTIFICIAL INTELLIGENCE

AI is commonly described as an ensemble of techniques to make machines mimic humans' intelligence. AI has been defined in many ways. However, it is out of the scope of this paper to give a formal definition of it. In the thematic diagram of Figure 10, AI it is certainly an emerging theme instead of a declining theme. Indeed, AI appeared in the DAR field only in recent years. Thus, it is still not very well-developed and central in the DAR literature as other themes. The AI field is composed of different sub-fields, and in this paper, we present the ML sub-field since it is the most frequent in the DAR research field.

ML methods are booming in recent years since they showed to improve both the speed and accuracy of the data analysis compared to traditional statistical tools [82]. ML is used to identify biotic and abiotic stress in crops. Crops stresses affect the yields, and their early detection leads to more efficient and effective interventions.

ML algorithms aim is to detect task-relevant patterns in the data without relying on (sometimes unjustifiable) assumptions. New information can be automatically obtained, and the algorithms learn in an automated way. This allows the patterns detection from data even if the underlying data model is unknown [82].

Data are usually arranged in matrices where each row is a sample composed of a certain number of features. A feature is a measurement of a characteristic of the data sample. For example, in the DAR field, features could be the spectral intensity along different wavelengths. Additionally, each element can be associated with a label. Common agricultural labels could be discrete like the plant's health state (e.g., healthy or diseased) or continuous like a crop unit yield.

ML methods are used for two main tasks: supervised and unsupervised learning. In supervised learning, a model is ''trained'' with labeled data and then used to predict labels of new unseen data. In unsupervised tasks, data are unlabeled, and the task is to find unknown patterns to obtain a new and more compact representation of the information [82].

The typical ML process is composed of many steps: data acquisition, data preparation, features selection, model selection, parameters selection, training, and scoring [83]. In particular, data acquisition, data preparation, and feature selection strongly influence the quality of the data. Data with a high level of noise, errors, outliers, biases may significantly reduce the model's prediction accuracy [84].

ML models are many, and the model choice is not unique and differs with tasks and crops. Typical ML algorithms are k-means clustering, Support Vector Machines (SVM), Decision Trees (DT), and (Artificial) Neural Networks ((A)NN). For detailed information about these methods, refer to the textbooks [85]–[87].

The authors of [28] made a review of ML applications in agriculture. In particular, they presented works grouped into four categories, and they made comparisons between them. These four categories are crop management, livestock management (known as Precision Livestock Farming (PLF)), water management, and soil management. Also, a good review of ML methods for crop yield and nitrogen status estimation is that of [84].

When ML algorithms are used to analyze images, the term Computer Vision is usually adopted. The authors of [83] reviewed agricultural Computer Vision applications related to disease detection, grain quality, and phenotyping. They also highlighted the main challenges for each of these tasks.

PLF is seeing a fast diffusion in recent years. PLF has ''the objective to create a management system based on continuous automatic real-time monitoring and control of the production/reproduction, animal health and welfare, and the environmental impact of livestock production'' [88]. Meat demand is increasing worldwide and has been projected growth of 40% from 2015 to 2030. Major concerns are related to the transmission of diseases to humans and the uncontrolled use of antibiotics leading to antibiotic resistance. The use of digital technologies in livestock production can be just as beneficial as in crop farming. PLF leads to a range of applications, from the monitoring of respiratory pathologies in intensive pig farms to the automatic detection of lameness problems in dairy cows. PLF applications are reviewed in [88].

Future trends regarding ML in agriculture foresee a greater integration of spatial, spectral, and temporal information with expert knowledge. Also, the integration of different ML techniques is expected to exploit different methods' strengths better. Finally, it is foreseeable the integration of multiple sources of data, from stationary (ground probes, weather stations) and mobile (UAV, UGV, satellites, tractors) platforms [84].

### 6) TRENDING TOPICS

The trending topics of the last three years (2017, 2018, 2019) are shown in Figure 11. The authors' keywords are vertically plotted on the corresponding year line and ordered by their frequency. For each year are plotted the twenty most frequent keywords. Each unique keyword in the dataset was associated with the median year based on the number of times that keyword has appeared in publications. In the following, the keywords shown in Figure 11 are *emphasized*.

The year 2017 has been characterized by *WSN* and *RS*. Relative to the latter topics, *hyperspectral imaging* received much attention. The *NDVI* showed to be the most common spectral index adopted and widely used for crop nitrogen mapping and *yield mapping* tasks. The use of spectral sensors in *RS* gave a renovated attention to the *SSM* topic.

The year 2018 has been dominated by the *IoT*, the *UAS*, and the concept of *CSA*. The latter has likely gained much attention by the Fridays for Future movements related to *climate change* and *sustainability* issues. Other trending topics of 2018 are the *PLF* and the *Big Data*.

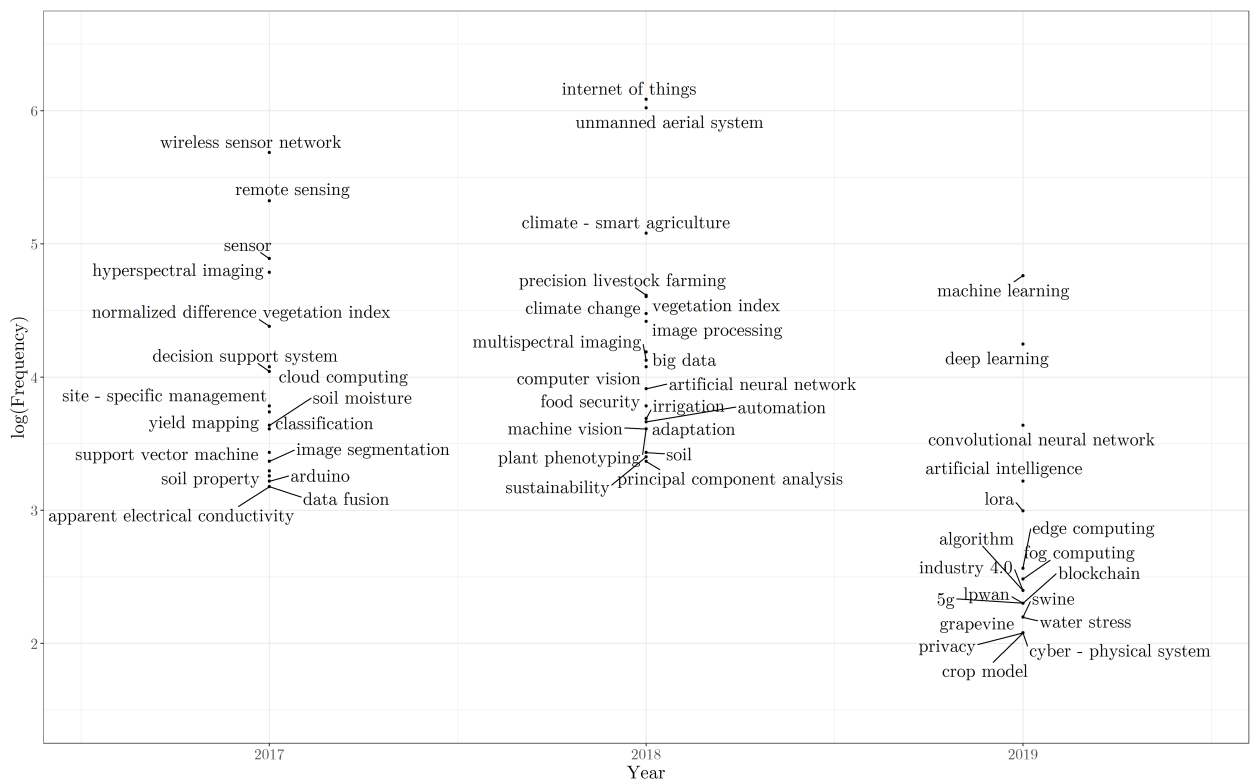

**FIGURE 11.** Trending topics of the years 2017, 2018, 2019. Each topic is assigned to the median year based on the number of publications containing it among the authors' keywords.

The year 2019 has been dominated by *AI* related topics like *ML*, *Deep Learning*, and *Convolutional Neural Network*. There has also been an interest on IoT related topics like the *edge and fog computing* concepts and IoT communications technologies like *LoRa*, the *5G* and *LPWAN*. Also, it is interesting to note the presence of the *Industry 4.0* term that is related to concept of Agriculture 4.0.

## V. CONCLUSION

In the last decade, digital technologies saw a rapid diffusion in the agricultural research field. Around the year 2012, there has been a steep increase in the number of publications related to the DAR themes. The yearly number of publications passed from about 250 in 2012 to more than 1250 publications in 2019. We showed how this growth is associated with the increased popularity of authors' keywords connected to the CSA, RS, IoT, UAS, and AI themes.

The concept of PA, which appeared in the literature since the '90s, is the basis on which other concepts like CSA and Agriculture 4.0 were built. The PA is the term that saw the more relevant increase in use (based on the number of publications) since 2012. This means that most authors have since now used the term PA to refer to the DAR. Digital technologies allow (or at least promise) to realize the PA concept in practice more efficiently and accurately than possible in the past.

After the 2007 IPCC, the need for climate-friendly agricultural practices became evident, and the new concept of CSA saw its birth. CSA expects agricultural practices to adapt to climate changes, reduce GHG emissions, and be sustainable. By examining the scientific literature, we realized that the CSA concept was most realized through digital technologies and referred to as Digital Agriculture or Smart Agriculture terms.

Finally, the popularity of the Industry 4.0 concept was mimicked in the agricultural sector, and in 2016, we saw the first appearance of the term Agriculture 4.0. Agriculture 4.0 ties together the concepts of PA and CSA, with a strong focus on digital technologies as tools to realize them. Also, Agriculture 4.0 focuses on the beyond-farm system, thus exploiting the information from the entire value chain, from the farmer to the consumer.

By clustering the authors' keywords reported on publications, we mapped five main topics in the DAR literature, that is, CSA, SSM, RS, IoT, and AI. The AI cluster classified as an emerging theme will likely become a motor theme in the following years, with high centrality and density in the DAR literature. Indeed, from the trending topics of the year 2019, we noted a strong presence of AI-related terms. Due to the always more incumbent necessity of a climate-friendly transition, it is also likely that the concept of CSA will gain centrality in the DAR literature. Instead, the SSM theme could get less attention in favor of a concept of per-plant management instead of homogeneous ground zones. The IoT and RS clusters that have been categorized as motor themes do not show any sign of getting less central in the near future. Based on the trending topics, the IoT field will likely experience a shift to the edge and fog computing paradigm. At the same time, the use of hyperspectral sensors will characterize RS in the following years.

This review article presented evidence of an ongoing DAR, what the DAR means, and where it is heading. However, some literature gaps are still present. Indeed, no articles provide an in-depth and broad analysis of Smart Agriculture, Digital Agriculture, and Agriculture 4.0 topics. Other sources of information, apart from the scientific literature, could be used too. Further investigation is needed to understand if the concepts of Smart Agriculture and Agriculture 4.0 are just a rebranding of PA or represent positive agricultural practices changes. Finally, more attention should be paid to develop objective metrics to prove that DAR technologies can meet the sustainability requirements and the extent to which digital technologies enable them.

## CONFLICTS OF INTEREST

The authors declare that they have no known competing financial interests or personal relationships that could have influenced the work reported in this paper.

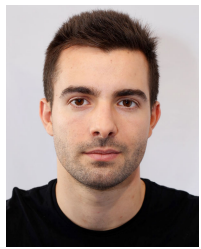

**RICCARDO BERTOGLIO** received the bachelor's and master's degrees in computer science and engineering from the Politecnico di Milano, where he is currently pursuing the Ph.D. degree in data analysis and decision sciences. His research interest includes the application of digital technologies in agriculture. In particular, he is specialized in robotics, AI, and computer vision research topics.

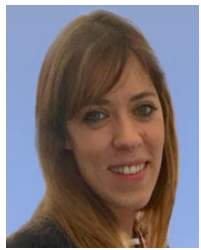

**CHIARA CORBO** received the bachelor's degree in business administration, the master's degree in marketing, and the Ph.D. degree in "agrisystem," the doctoral school for the agro-food system from the Università Cattolica del Sacro Cuore. She was a Researcher in innovation and sustainability in the agri-food sector with more than ten years of experience. She worked at various research institutes and consulting companies in Italy and abroad, dealing mainly with quality, traceability, sustainability, and food production safety. She is currently a Researcher at the Digital Innovation Observatory, School of Management, Politecnico di Milano, within the Smart AgriFood Observatory, where she studies the impacts of digital innovation on agrifood supply chains. Her research interests include regard digital innovation in the agri-food sector, precision farming, food safety and traceability, and food sustainability. On these topics, she is the coauthor of several academic publications and popular articles.

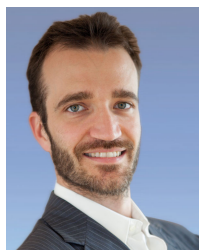

**FILIPPO M. RENGA** is currently the Co-Founder of the Digital Innovation Observatories, School of Management, Politecnico di Milano, where he started and coordinated the mobile (mobile and app economy, marketing and service, payment and commerce, banking, and enterprise), digital innovation in tourism, fintech and insurtech, and smart agrifood observatories, where he is also the Coordinator of the paths of excellence at the Cremona Campus. He is also the Co-Founder of five startups with a turnover of more than €35 million to date. He is the coauthor of various academic articles on mobile services.

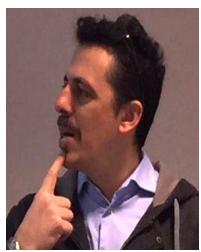

**MATTEO MATTEUCCI** (Member, IEEE) received the Laurea degree in computer engineering from the Politecnico di Milano, Milan, Italy, in 1999, the Master of Science degree in knowledge discovery and data mining from Carnegie Mellon University, Pittsburgh, PA, USA, in 2002, and the Ph.D. degree in computer engineering and automation from the Politecnico di Milano, in 2003. He is currently a Full Professor at the Dipartimento di Elettronica Informazione e Bioingegneria, Politecnico di Milano. He has coauthored more than 50 (peer-reviewed) articles in international journals, 25 papers in international books, and more than 150 (peer-reviewed) contributions to international conferences and workshops. He has been the Principal Investigator in national and international funded research projects on machine learning, autonomous robots, sensor fusion, and autonomous and intelligent systems benchmarking. His main research interests include pattern recognition, machine learning, machine perception, robotics, computer vision, and signal processing. He is also interested in developing, evaluating, and applying, in a practical way, techniques for adaptation and learning to autonomous systems interacting with the physical world.

○ ○ ○